
%
%
\input phyzzx.tex
\magnification=\magstephalf

\font\largemath = cmti10 scaled \magstep2
%
\def\Ge2{\Ge^2}
\def\GL2{\GL^2}
\def\Ga{\alpha}
\def\Gb{\beta}

\def\Ge{\epsilon}

\def\Gg{\gamma}

\def\GL{\Lambda}




%

%
%

\def\MKO{\Bigl[}
\def\MKC{\Bigr]}

\def\MCO{\Bigl\{}
\def\MCC{\Bigr\}}

%


\REF \paris{G.Parisi and Y.Wu, Sci. Sin. {\bf 24} (1981)483}
\REF \damgf{For a review, P.H.Damgaad and H.H\"uffel, Phys. Rep.
{\bf 152} (1987)227}
\REF \gh{M.Namiki, I.Ohba, K.Okana and Y.Yamanaka, Prog. Theor. Phys. {\bf
69} (1983)\break 1580}
\REF \nakgh{A.Nakamura, Prog. Theor. Phys. {\bf 86} (1991)925}
\REF \namoo{M.Namiki, I.Ohba and K.Okano, Prog. Theor. Phys. {\bf 72} (1984)
350}
\REF \ito{K.Ito, Proc. Imp. Acad. {\bf 20} (1944)519}

\REF \gra{R.Graham, Phys. Lett. {\bf 109A} (1985)209;\hfill\break
R.Mochizuki, Mod. Phys. Lett. {\bf A5} (1990)2335;\hfill\break
H.Rumpf, Phys. Rev. {\bf D33} (1986)942}

\REF \riuji{R.Mochizuki, Prog. Theor. Phys. {\bf 85} (1991)407}

\REF \nlsm{J.Zinn-Justin, Nucl. Phys. {\bf B275[FS17]} (1986) 135;
\hfill\break N.Nakazawa and D.Ennyu, Nucl. Phys. {\bf B305[FS23]} (1988)
516}
\REF \fadd{L.D.Faddeev,Teoret. i Mat. Piz. {\bf 1} (1969)3; \hfill\break
P.Senjanovich, Ann. of Phys. {\bf 100} (1976)227;\hfill\break
T.Maskawa and H.Nakajima, Prog. Theor. Phys. {\bf 56} (1976)1295}


\REF \arn{L.Arnold, Stochastic Differential Equations (Wiley-Intersciense,
New York,\break 1974)}
\REF \stra{T.L.Stratonovich, Conditional Markov processes and their
application to the theory of optimal control (Elsevier, New York, 1968)}
\REF \waseda{H.Kawara, M.Namiki, H.Okamoto and S.Tanaka,
Prog. Theor. Phys. {\bf 84} (1990)749; \hfill\break
N.Komoike, Prog. Theor. Phys. {\bf 86} (1991)575}

\textfont4=\largemath
\mathchardef\lS="0453
\footline={\hfill}
\unnumberedchapters
\date={}
%
\pubnum{CHIBA-EP-57}
\date{revised, September 1992}
\titlepage
\title{Treatment of Constraints in Stochastic Quantization Method
and Covariantized Langevin Equation}
\author{Kenji Ikegami}
\address{Graduate school of Science and Technology, \break
          Chiba University,  \break
          1-33 Yayoi-cho, Inageku, Chiba 263, Japan}
\author{Tadahiko Kimura and Riuji Mochizuki}
\address{Department of Physics, Faculty of Science, \break
        Chiba University,  \break
          1-33 Yayoi-cho, Inageku, Chiba 263, Japan}
\baselineskip = 12pt
\abstract{
We study the treatment of the constraints in stochastic quantization
method. We improve the treatment of the stochastic consistency condition
proposed by\break Namiki et al. by suitably taking account of the Ito calculus.
Then we obtain an improved Langevin equation and
the Fokker-Planck equation which naturally leads to the correct path
integral quantization of the constrained system as the stochastic
equilibrium state. This treatment is applied to $O(N)$ non-linear $\sigma$
model and it is shown that singular terms appearing in the improved
Langevin equation cancel out the $\delta^n(0)$ divergences in one loop
order. We also ascertain that the above Langevin equation, rewritten in terms
of
independent variablesis, actually equivalent to the one in the
general-coordinate-transformation-covariant and vielbein-rotation-invariant
formalism.
 }
\footnote{}{$*$  To be published in Nucl. Phys. B}
\def\underset#1\to#2{\mathop{#2}\limits_{#1}}

\baselineskip = 18pt
\footline={\hfill-- \folio\ --\hfill}
\section{1. Introduction}

 Stochastic quantization method (SQM) was first proposed by Parisi and
Wu.\refmark\paris\refmark\damgf They showed that the method could be
applied to gauge theory without the gauge fixing procedure. That is, in
SQM it is not necessary to introduce the Faddeev-Popov ghost fields.
Nevertheless the method produces the same contributions as those due to
ghost fields, which was perturbatively confirmed first for Yang-Mills
field\refmark\gh
and recently for non-Abelian anti-symmetric tensor field.\refmark\nakgh

 How to handle the constrained system in SQM was discussed by
Namiki et al. in ref.5. They constructed Langevin equation for the system
under the holonomic constraints by imposing the stochastic consistency
condition. In the path integral quantization method the constraints introduce
a determinant factor into path integral measure, which requires that in SQM
Langevin equation for the constrained system is constructed so that
the equilibrium Fokker-Planck distribution derived from the Langevin
equation has the same determinant factor. They showed
that the equilibrium distribution coincided with the path integral
distribution.
Nevertheless they had to use the 5-dimensional stochastic
path integral representation of the transition probability distribution of
stochastic process and could not derive Fokker-Planck equation directly from
Langevin equation, because they did not take account of Ito
calculus\refmark\ito
in their treatment of stochastic consistency
condition.
%
One of the main purposes of this paper is to improve Langevin equation of
ref.5 by suitably taking account of Ito calculus, and to show that
the improved Langevin equation leads to the Fokker-Planck equation which
directly gives the correct path integral representation as the
equilibrium distribution.
Our improvement introduces some singular terms proportional to
$\delta(0)$ in the Langevin equation. So, the Langevin equation
of ref.5 is correct
in the dimensional regularization scheme, but not in other regularization
schemes
and the general path integral distribution with constraints
cannot be directly obtained as the equilibrium Fokker-Planck distribution.

 On the other hand, if the constraints are solved explicitly and the
system is described in terms of the independent variables only, the action of
the system proves to have generally field-dependent metric. In this
case we must apply the general-coordinate-transformation (GCT)-covariant
and vielbein-rotation (VR)-invariant \break Langevin equation to
the system.\refmark\gra
It was
not clear whether the latter Langevin equation is equivalent to
the above Langevin equation for the constrained system. This point will be
clarified in this paper.

 Next we discuss about O(N) non-linear $\sigma$ model as an example
of such constrained system. The model was studied by many
authors.\refmark\nlsm   We apply the above two methods to the model and
clarify the role of singular terms introduced into the Langevin equation
with the stochastic perturbation theory. It will be shown that in both
these methods these singular terms are necessary to cancell the $\delta^n(0)$
divergences appearing in one-loop expansion.

 This paper is organized as follows. In section 2 we study
the improved treatment of the constraints in SQM and show directly that
the equilibrium Fokker-Planck distribution coincides with the path
integral distribution. In section 3 it is shown that the improved Langevin
equation for the constrained system is equivalent to the GCT-covariant
and VR-invariant Langevin equation. In section 4 we apply both the
improved Langevin equation for the constrained system and the GCT-covariant
and VR-invariant Langevin equation to O(N) non-linear $\sigma$ model and
examine the cancellation of $\delta^n(0)$ divergences. In section 5 we give
conclusion and summary. In Appendix, we ascertain that an assumption,
which is introduced in section 3, is satisfied
at least in O(N) non-linear $\sigma$ model.

\section{2.Constrained system in SQM}
 In this paper we consider the system with variables
$q_i (x) (i=1,2,\cdots,N)$, regular Lagrangian
$L(q_i,\partial_\mu q_i)\ (\mu =1,2,\cdots ,n)$ and a set of
constraints
  $$
  F_a(q_i)=0 ,\ \ \ \ \ \ \ \ \    (a=1,2,\cdots,M;N>M). \eqno (2.1)
 $$

  In the path integral quantization method the
transition amplitude is given by\refmark\namoo\refmark\fadd
  $$
     \langle f \mid i \rangle = \int Dq DJ \ \sqrt {det
       D_{ab}} \ exp[-\int d^nx(L(q,\partial q) -J_a F_a)].\eqno(2.2)
      $$
      $$
      D_{ab} \equiv {\partial F_a \over \partial q_i}{\partial F_b \over
    \partial q_i}.\eqno (2.3)
  $$

Following the method ``time by time constraint'' proposed in ref.5,
we quantize the above singular system in SQM. The treatment of the
stochastic consistency condition is improved by taking account of Ito
calculus. It will be shown that the improvement is essential for the
acquisition
of the Fokker-Planck equation which directly leads to the path integral
representation as the equilibrium state.

  For the above system, Langevin equation is\refmark\namoo
 $$
   dq_i (t) \equiv q_i(t + dt) - q_i(t)
    = - {\delta S \over \delta q_i}dt -
    {\partial F_a \over \partial q_i}\lambda_a dt + dW_i,\eqno(2.4)
    $$
  $$
  S \equiv \int d^nx L,\eqno(2.5)
 $$
where $\lambda_a$ is Lagrange multiplier and $dW_i$ is defined as
 $$
   dW_i(t) \equiv W_i(t+dt)-W_i(t), \eqno (2.6)
      $$
        $$
           W_i(t) \equiv \int^t dt'\eta_i(t'), \eqno (2.7)
        $$
       $$
       \langle dW_i(x,t)dW_j(x',t)\rangle = 2\delta_{ij}\delta(x-x') dt,
      \eqno(2.8)
     $$
   $$
    \langle \eta_i(x,t)\eta_j(x',t') \rangle =
    2\delta_{ij} \delta^n(x-x') \delta(t-t').\eqno(2.9)
 $$
and called Wiener process. From (2.8) we may regard $dW$ as
order $\sqrt{dt}$. Lagrange multiplier $\lambda_a$ is determined by
the stochastic consistency condition\refmark\namoo
 $$
   \dot F_a(q(t)) = 0,\eqno(2.10)
 $$
where dot denotes fictitious-time derivative. Besides, we demand the
initial condition
   $$
     F_a(q(t=t_0))=0,\eqno(2.11)
    $$
in order to have $F=0$ at any {\it t}.
In SQM, for fictitious-time derivative of any function $f(q)$
\refmark\ito\refmark\arn
 $$
   {d\over dt}f(q(t)) \not= {\partial f\over \partial
   q}\dot{q},\eqno(2.12)
  $$
unless we use Stratonovich calculus.\refmark\stra In ref.5 the above fact is
not taken into account.
(Their Langevin equation is correct only in the dimensional
regularization scheme. In other
regularization schemes, the Langevin equation is incorrect and
the Fokker-Planck distribution does not coincide with the path integral
distribution.) %
If $\lambda dt$ in eq.(2.4) does not contain terms of order $\sqrt{dt}$,
$dq=dW$ to order $\sqrt{dt}$. From (2.4) we
get
 $$
  dF_a={\partial F_a\over\partial q_i}\lbrace
       -({\delta S\over\delta q_i}
    + \lambda_a {\partial F_a \over \partial q_i})dt + dW_i \rbrace
   + {1\over 2}(1-2b){\partial^2 F_a \over \partial q_i\partial q_j}
  dW_i dW_j + O((\sqrt{dt})^3),\eqno(2.13)
 $$
to order $dt$. Here we use the generalized Ito formula
\refmark\riuji and the product of Wiener process $dW$ and any function
$f(q)$ is defined as
 $$
   \lbrace f(q)\ dW \rbrace (t) \equiv \lbrace bf(q(t+dt)) +
     (1-b)f(q(t)) \rbrace dW(t),
    $$
    $$
   0 \leq b \leq 1, \eqno(2.14)
 $$
where $b={1\over 2}$ corresponds to Stratonovich calculus and
$b=0$ to Ito calculus. Requiring $dF_a = 0$,
we get
  $$
   \lambda_a dt = D^{-1}_{ab} \lbrace{\partial F_b \over \partial q_i}
      (-{\delta S \over \delta q_i}dt + dW_i) +{1 \over 2}(1-2b)
        {\partial^2 F_a \over \partial q_i \partial q_j}dW_i dW_j
          \rbrace,\eqno(2.15)
         $$
         $$
       D_{ab} \equiv {\partial F_a \over \partial q_i}{\partial F_b
    \over \partial q_i} .\eqno(2.16)
  $$
Here $\lambda_a dt$ contains
$D^{-1}_{ab}{\partial F_b \over \partial q_i}dW_i$
of order $\sqrt{dt}$, which is inconsistent with
the above assumption that $\lambda_a dt$ does not contain terms of order
$\sqrt{dt}$. So we assume alternatively that
the $\lambda_a dt$ term contains terms of order $\sqrt{dt}$ like
eq.(2.15). Then, from (2.4)
 $$
   dq_i = dW_i - {\partial F_a \over \partial q_i}\lambda_a dt\ \ \
\ \ \ \ \ \ \ \ \ \ \ \ \
   $$
    $$
         = dW_i - {\partial F_a \over \partial q_i} D^{-1}_{ab}
    {\partial F_b \over \partial q_j} dW_j,\eqno(2.17)
  $$
to order $\sqrt{dt}$. From (2.17) the modified expression
of $dF$ is
 $$
   dF_a = {\partial F_a \over \partial q_i}\lbrace
     (-{\delta S \over \delta q_i} - \lambda_b{\partial F_b \over
      \partial q_i})dt + dW_i \rbrace
          +{1\over 2}(1-2b){\partial^2 F_a \over \partial q_i
           \partial q_j} K_{ik} dW_k K_{jl} dW_l,\eqno(2.13')
         $$
        $$
      K_{ij} \equiv \delta_{ij} - R_{ij},\ \ \ \ \ \ \
     R_{ij} \equiv {\partial F_a \over \partial q_i}
    D^{-1}_{ab} {\partial F_b \over \partial q_j},\eqno(2.18)
  $$
to order $dt$. Here $R_{ij},K_{ij}$ are projection
operators, vertical each other. The final term in RHS of $(2.13')$
does not exist in ref.5. From the consistency condition
(2.10) the correct expression of $\lambda dt$ becomes
  $$
     \lambda_a dt = D^{-1}_{ab}\lbrace{\partial F_b \over
        \partial q_i}(-{\delta S \over \delta q_i}dt + dW_i)
          + {1\over 2} (1-2b) { \partial^2 F_b \over \partial q_i
         \partial q_j} K_{ik} dW_k K_{jl} dW_l\rbrace.\eqno(2.15')
  $$
The above expression is surely {\it correct}, because, with the help of
$(2.15')$, $dq$ has the same terms as (2.17) to order $\sqrt{dt}$. From
(2.8) and $(2.15')$, Langevin equation (2.4) becomes
 $$
   \dot{q}_i = K_{ij}(-{\delta S \over \delta q_j} + \eta_j)
      -(1-2b){\partial F_a \over \partial q_i}D^{-1}_{ab}
        {\partial^2 F_b\over \partial q_k \partial q_l}K_{kl}
         \delta^n(0) ,\eqno(2.19)
       $$

 If we multiply (2.19) by ${\delta F_a \over \delta q_i}$, we obtain
the expression $(2.13')$ or (2.10). Therefore, the consistency condition (2.10)
is embedded in eq.(2.19) and eq.(2.19) means {\it N-M} independent
differential equations. In ref.5 Langevin equation did not have the
singular term in (2.19). The same equilibrium Fokker-Planck distribution as
(2.2) could not be derived directly from the Langevin equation of ref.5, while,
due to the singular term, we can derive the correct Fokker-Planck equation
directly from Langevin equation (2.19) as shown below.

 In order to construct the Fokker-Planck equation, we
introduce the expectation value of fictitious-time derivative of arbitrary
function $g(q)$
  $$
     \langle \dot g(q(t)) \rangle \equiv \int Dq \ g(q)
         \dot P(q,t),\eqno(2.20)
  $$
where $P(q,t)$ is the transition probability distribution. Using
integral by parts and generalized Ito formula, we obtain
the Fokker-Planck equation
 $$
   \dot P(q,t) = {\delta \over \delta q_i}K_{ij}
            \lbrace {\delta S \over \delta q_j}
              -{\partial^2 F_a \over \partial q_j \partial q_l}
                D^{-1}_{ab}{\partial F_b \over \partial q_l}
          \delta^n(0) +{\delta \over \delta q_j}\rbrace
        P(q,t).\eqno(2.21)
   $$
due to the singular term in (2.19). Eq.(2.21) cannot be derived from
the Langevin equation of ref.5. The probability distribution can include
any function $f(F_a)$ because ${\delta F_a \over \delta q_i}$ is vertical to
the projection operater $K_{ij}$. In the equilibrium limit
$t\rightarrow\infty$, the probability distibution must satisfy  $\dot P =
0$. In the limit the equation has a solution
  $$
    P(q) =\int Dv_a \sqrt{detD_{ab}} exp(-S-\int d^nx\ v_aF_a)
     ,\eqno(2.22)
 $$
where we chose $\int Dv exp(-v_aF_a)$ as $f(F_a)$ in accordance with the
initial condition (2.11) in Minkowski space. The above equilibrium
probability distribution coincides with eq.(2.2).
However, it is strange that the Langevin equation has
divergent term of $\delta^n(0)$. In section 4 we examine O(N) non-linear
$\sigma$ model
as an example of the system under constraint (2.1). In the same example
we also show perturbatively that the term proportional to $\delta^n(0)$ in
eq.(2.19) is needed.

\section{3.Equivalence to GCT-covariant \hfill\break
$\cdots \cdots \cdots$ and VR-invariant Langevin equation }

 In this section we shall show that the improved Langevin equation (2.19)
is actually equivalent to the Langevin equation in the GCT-covariant and
VR-invariant formalism where the constraint (2.1) is explicitly solved and
the equation is expressed in terms of independent variables only.

In general
the system may have field-dependent metric (or kernel) $G_{AB}(q)$ after the
constraint is solved. According to ref.7, a system with field-dependent metric
is described by the GCT-covariant and VR-invariant Langevin equation
  $$
     dq^A=X^Adt+E_m^{\ A}dW^m,\eqno(3.1)
   $$
where
  $$
       X^A\equiv -G^{AB}{\delta S\over\delta q^B}+{1\over\sqrt{G}}
        {\delta\over\delta q^B}(\sqrt{G}G^{AB})-2b{\delta E_m^{\ A}
       \over\delta q^B}E_n^{\ B}\delta^{mn},
     $$
       $$
         G^{AB} = E_m^{\ A} E_n^{\ B} \delta^{mn}, \ \ \ \ A,B,m,n =
              (1,2,\cdots,N-M;x).\eqno(3.2)
  $$
$G^{AB}$ is the inverse of metric $G_{AB}$, the summation
with respect to B includes space-time integration and $dW^m$ is Wiener process
defined in section 2. GCT-covariance and VR-invariance mean
that Langevin equation is transformed covariantly
under general coordinate transformation
$q\rightarrow q'=f(q)$ and is invariant under vielbein rotation
$E^A_{\ m}\rightarrow E^A_{\ n}\Lambda^n_{\ m}$. $dq^A$ and $E_m^{\ A}dW^m$ are
not transformed covariantly and two extra terms in (3.2) are
required to be GCT-covariant and VR-invariant.

 In order to decompose variables into constraint variables and
independent ones, we introduce a new set of variables $\{ Q^{\mu}\}
(\mu =1,2,\cdots,N)$.\refmark\namoo  $Q^{\mu}$'s are
expressed in terms of $q_i$'s $(i=1,2,\cdots,N)$ as
  $$
        \delta Q^{\mu}=e^{\mu}_{\ i} \delta q_i,\ \ \ or \ \ \
        {\partial Q^{\mu} \over \partial q_i} = e^{\mu}_{\ i},\eqno(3.3)
  $$
where $e^{\mu}_{\ i}$ is vielbein field defined as follows. First
$e^a_{\ i}$ and $\ e_{a,i}\ (a=N-M+1,\cdots,N)$ are defined as
  $$
     e^a_{\ i}= {\partial F_a \over \partial q_i},\ \ \ \
     e_{a,i}=D^{-1}_{ab}{\partial F_b \over \partial q_i},\eqno(3.4)
  $$
i.e. $Q^a=F_a$. Then, $e^{A}_{\ }i\ (A=1,2,\cdots,N-M)$ is chosen so
as to satisfy
  $$
     e^A_{\ i} e_{a,i} = 0,\eqno(3.5)
  $$
and its inverse $e_{A,i}$ is defined as
  $$
      e_{A,i} = (g^{-1})_{AB}e^B_{\ i},\ \ \ \
      g^{AB} \equiv e^A_{\ i}e^B_{\ i},\eqno(3.6)
  $$
where we assume that $g^{AB}$ is non-singular. From the above definition
it turns out that $e^{\mu}_{\ i}$ and $\ e_{\mu ,i}$ satisfy the following
relations
  $$
     e^{\mu}_{\ i} e_{\nu ,i} = \delta^{\mu }_{\nu },\ \ \ \ \
      e^{\mu}_{\ i} e_{\mu ,j} = \delta_{ij},\eqno(3.7)
       $$
        $$
          e^A_{\ i} e^a_{\ i} = e^A_{\ i} e_{a,i} = e_{A,i} e^a_{\ i}
           = e_{A,i} e_{a,i} =0,\eqno(3.8)
          $$
           $$
           e^A_{\ i} e_{A,j} = K_{ij},\ \ \ \ \ e^a_{\ i} e_{a,j} =
            R_{ij},\eqno(3.9)
          $$
          $$
         K_{ij} e^a_{\ j} = K_{ij} e_{a,j} = 0,\ \ \ \ \
       R_{ij} e^A_{\ j} = R_{ij} e_{A,j} = 0.\eqno(3.10)
     $$
    $$
   det(e^{\mu}_{\ i}) \not= 0.\eqno(3.11)
  $$
{}From (3.11) the manifold spanned by $q_i$'s is identical with the one
by $Q^{\mu}$'s. With the help of the same discussion as made about
eq.$(2.13')$, $dQ^{\mu}$ is written as follows
  $$
   \eqalign { dQ^{\mu}\equiv Q^{\mu}(t+dt)-Q^{\mu}(t)
         &= {\partial Q^{\mu} \over \partial q_i} dq_i +
     (1-2b){\partial^2 Q^{\mu} \over \partial q_i \partial
                     q_j}K_{ij}dt\delta^n(0),\cr
         &=e^{\mu}_{\ i} dq_i +
     (1-2b) {\partial e^{\mu}_{\ i} \over \partial q_j}
           e^A_{\ }i e_{A,j}dt\delta^n(0).}\eqno(3.12)
       $$
Then, from ($2.13'$) constraint variables $Q^a$'s satisfy
Langevin equation
  $$\eqalign{
    dQ^a
     &= e^a_{\ i} \lbrace K_{ij}(-{\delta S \over \delta q_{\ j}}dt+ dW_j)
         - (1-2b) {\partial F_a \over \partial q_i} D^{-1}_{ab}
           {\partial^2 F_b \over \partial q_j \partial q_l}K_{jl}dt
          \delta^n(0) \rbrace \cr
     &\ \ \ \ \ \ \ \ \ \ \ \ \ \ \ \ \
    + (1-2b) {\partial e^{a}_{\ i} \over \partial q_j} e^A_{\ i}e_{A,j}dt
      \delta^n(0),\cr
     &= 0,}\eqno(3.13)
  $$
to order {\it dt}. From (3.13) and the initial condition (2.11),
constraint variables $Q^a$'s are zero for all {\it t}. As for the
independent variables $Q^A$'s we get
  $$
   \eqalign {
      dQ^A
      &= e^A_{\ i} \lbrace K_{ij}(-{\delta S \over \delta q_j }dt + dW_j)
    - (1-2b) {\partial F_a \over \partial q_i} D^{-1}_{ab}
      {\partial^2 F_b \over \partial q_j \partial q_l}K_{jl}dt
         \delta^n(0) \rbrace \cr
      &\ \ \ \ \ \ \ \ \ \ \ \ \ \ \ \ \
    + (1-2b) {\partial e^{A}_{\ i} \over \partial q_j} e^B_{\ i} e_{B,j}dt
         \delta^n(0),\cr
      &= -g^{AB}{\delta S \over \delta Q^B}dt +
       (1-2b){\partial e^{A}_{\ i} \over \partial Q^B}e^B_{\ i}dt
            \delta^n(0)  + {e^A_{\ i} dW^i}.}\eqno(3.14)
     $$
The above Langevin equation is not invariant under vielbein rotation
$e^A_{\ i}\rightarrow e^A_{\ j}\Lambda^j_{\ i}(Q)$ because Wiener
process $dW^i$ is defined in a manifold spanned by original variables $q_i$'s
and we must not consider the rotation in the manifold. If we
perform field-dependent rotation in the manifold, $\Lambda^i_{\ j}dW^j$
is not Wiener process, i.e.$\langle \Lambda^i_{\ j}dW^j \rangle \not= 0$.
In order to reduce (3.14) to the form of (3.1) we decompose the
vielbein as follows:
  $$
     e^A_{\ i} =E^A_{\ I}(Q) \epsilon^I_{\ i}(Q),\ \ \ \
           I=(1,2,\cdots,N-M),\eqno(3.15)
    $$
       $$
          \epsilon^I_{\ i} \epsilon^J_{\ i} = \delta^{IJ},\ \ \ \
          e^A_{\ i} e^B_{\ i} = E^A_{\ I} E^B_{\ J} \delta^{IJ} =
             g^{AB},\eqno(3.16)
          $$
and define $dW^I$ by
            $$
               dW^I \equiv \epsilon^I_{\ i} dW^i -
           2b{\partial \epsilon^{I}_{\ i} \over \partial Q^B}
           E^B_{\ J}\epsilon^{J}_{\ i}\delta^n(0).
             \eqno(3.17)
         $$
Then, we obtain
        $$
       \langle dW^I \rangle = 0,\ \ \ \ \
    \langle dW^I(x) dW^J(y) \rangle = 2\delta^{IJ}dt \delta^n(x-y)
              .\eqno(3.18)
   $$
$dW^I$ is desirable Wiener process and with $dW^I$ Langevin
equation is written as
   $$
     dQ^A = -g^{AB}{\delta S \over \delta Q^B}dt +
    {\partial \over \partial Q^B} (E^{A}_{\ I}\epsilon^I_{\ i})
           E^B_{\ J} \epsilon^J_{\ i}dt \delta^n(0)
        -2b{\partial E^{A}_{\ I} \over \partial Q^B}E^B_{\ J}\delta^{IJ}dt
        \delta^n(0) + E^A_{\ I} dW^I.\eqno(3.19)
   $$
{}From now on we regard $E^A_{\ I}$ as vielbein corresponding to $E^A_{\ m}$ in
(3.1). In fact the above Langevin equation is invariant under vielbein rotation
  $$
        E^A_{\ I} \rightarrow E^A_{\ J} \Lambda^J_{\ I}.\eqno(3.20)
  $$
Langevin equation (3.19) is a little different from GCT-covariant and
VR-invariant Langevin equation (3.1). Here we assume
   $$
      e^A_{\ i}\nabla_B e^B_{\ i} \equiv {1 \over \sqrt{g}}{\partial \over
\partial
                 Q^B}(\sqrt{g} e^B_{\ i})\ e^A_{\ i} =0,\eqno(3.21)
   $$
where $\nabla_B$ is covariant derivative in Riemanian manifold spanned by
$Q^A$.
Eq.(3.21) is usually presumed, because $g^{AB}$ satisfies
the metric condition $\nabla_A g^{BC}=0$ and the latter condition makes
eq.(3.21) naturally understandable. We ascertain in Appendix
that the above assumption is satisfied in O(N) non-linear $\sigma$ model.
With (3.21), Langevin equation
(3.19) is reduced to
  $$
     dQ^A = -g^{AB}{\delta S\over\delta Q^B}dt+{1\over\sqrt{g}}
        {\partial\over\partial Q^B}(\sqrt{g}g^{AB})\delta^n(0)dt
             -2b{\partial E^A_{\ I} \over\partial Q^B}E^B_{\ J}
                  \delta^{IJ}\delta^n(0)dt+E^A_{\ I}dW^I, \eqno(3.22)
           $$
        $$
      g \equiv det(g_{AB}). \eqno(3.23)
  $$
Covariant derivative in (3.21) does not include spin
connection because, as mentioned above, the rotation must not be considered
in the manifold spanned by $q_i$'s. Eq.(3.22) is equivalent to
GCT-covariant and VR-invariant Langevin equation (3.1).
Thus it is ascertained that Langevin equation (2.19) for the constrained
system is equivalent to GCT-covariant and VR-invariant Langevin equation
(3.1). That Langevin equation also has divergent term including $\delta^n(0)$
and, in the next section, we apply the Langevin equation to O(N) non-linear
$\sigma$ model and see the cancellation of $\delta^n(0)$ divergences.

\section{4. O(N) non-linear $\sigma$ model}

 O(N) non-linear $\sigma$ model is defined by action
 $$
    S = {1\over 2}\int d^nx \partial_{\mu}\Phi_i
      \partial_{\mu}\Phi_i,\ \ \ \ \
      (\mu =1,2,\cdots ,n; i=1,2,\cdots ,N),\eqno(4.1)
         $$
and constraint
   $$
       F = \Phi_i\Phi_i - {1\over \alpha} =0,\eqno(4.2)
 $$
where $\alpha$ is constant. The constraint is an example of (2.1). Applying
Langevin equation (2.19) to the model, we obtain
  $$
    \dot{\Phi}_i = (\delta_{ij} - \alpha \Phi_i
       \Phi_j)(\partial^2\Phi_j + \eta_j) -
     (1-2b)\alpha(N-1)\Phi_i\delta^n(0) .\eqno(4.3)
  $$
The above equation includes (2.19)-type constraint (4.2) and means {\it N-1}
independent equations. From eq.(4.2) $\Phi_i$ has non-zero vacuum expectation
value. We shift the
field
  $$
     \phi_i \equiv \Phi_i-<\Phi_i>,\eqno(4.4)
       $$
       $$
     v_i \equiv {1\over \sqrt\alpha}<\Phi_i>,\ \ \ \ \ v_iv_i=1,\eqno(4.5)
  $$
and with the shifted field $\phi$ Langevin equation is written
 $$
      \dot{\phi}_i = \lbrace K_{ij} - \sqrt{\alpha}(v_i\phi_j +
         v_j\phi_i) - \alpha \phi_i \phi_j \rbrace \lbrace \partial^2
           \phi_j + \eta_j \rbrace
         $$
           $$
       - (1-2b)\alpha(N-1)({1 \over \sqrt{\alpha}}v_i + \phi_i)\delta^n(0)
        ,\eqno(4.6)
      $$
     $$
   K_{ij} \equiv \delta_{ij} - v_i v_j.\eqno(4.7)
  $$
Going to momentum space and integrating eq.(4.6) with respect to {\it t}, we
get
\eject
  $$
     \phi_i(k,t) = \int^t d\tau  G_{ij}(k,t-\tau) \lbrack \ K_{jl}
       \ \eta_{l}(k,\tau) + I_j(k,\tau) + J_j(k,\tau)
         $$
           $$
            - (1-2b) \alpha (N-1)\lbrace{v_j \over \sqrt{\alpha}} +
             \phi_j(k,\tau) \rbrace \delta^n(0) \rbrack,\eqno(4.8)
            $$
             $$
             G_{ij}(k,t) \equiv \lbrack exp \lbrace -K\ k^2t \rbrace
               \rbrack_{ij} = exp(-k^2t)K_{ij} +
           (\delta_{ij} - K_{ij}),\eqno(4.9)
          $$

           $$
             I_i(k,t) \equiv - \sqrt{\alpha} \int {d^np d^nq \over
                  (2\pi)^n} \delta^n(k-p-q)(v_i \delta_{lm} + v_m
                     \delta_{il}) \phi_l(p,t) \lbrace -q^2 \phi_m(q,t) +
                       \eta_m(q,t) \rbrace ,\eqno(4.10)
       $$
      $$
     J_i(k,t) \equiv - \alpha \int {d^np d^nq d^nr \over (2\pi)^{2n}}
             \delta^n(k-p-q-r) \phi_i(p,t) \phi_j(q,t) \lbrace -r^2
          \phi_j(r,t) + \eta_j(r,t) \rbrace,\eqno(4.11)
  $$
After solving the above equation by iteration, we express the result
graphically
  $$
                  Fig.\ 1
  $$
where we denote $\eta$ by a cross or an encircled cross, $G$ a line
and $\alpha \delta(0)$ a bullet, respectively. We calculated the
one-loop corrections of the two-point function and obtained six
$\delta(0)$-divergent diagrams.
  $$
                  Fig.\ 2
  $$
Each of the six diagrams contributes respectively
  $$
     2\alpha N \delta^n(k+k') K_{ij} {1 \over (k^2+k'^2)^2}\int
       d^np,\leqno(2a)
    $$
      $$
      -4\alpha N \delta^n(k+k') \theta (0) K_{ij} {1 \over (k^2+k'^2)^2}
         \int d^np,\leqno(2b)
        $$
          $$
         2\alpha v^2 \delta^n(k+k') K_{ij} {1 \over (k^2+k'^2)^2}
       \int d^np,\leqno(2c)
         $$
         $$
        -4\alpha v^2 \delta^n(k+k') K_{ij} {1 \over (k^2+k'^2)^2}\int
       d^np,\leqno(2d)
       $$
      $$
     4\alpha v^2 \delta^n(k+k') \theta(0) K_{ij} {1 \over (k^2+k'^2)^2}\int
       d^np,\leqno(2e)
    $$
   $$
  -2\alpha(1-2b)(N-1)(2\pi)^n \delta^n(0) \delta^n (k+k') K_{ij}
 {1\over(k^2+k'^2)^2}.\leqno(2f)
  $$
The contributions from all the $\delta(0)$-divergent diagrams {\it cancel out}
if we put $\theta(0)=b$.\refmark\waseda It turns out that
$\delta^n(0)$-divergent term of eq.(2.19) or eq.(4.3) is necessary to the
cancellation of the $\delta(0)$ divergences in one-loop order.

Now we apply Langevin equation (3.1) to the model. As discussed in
section 3, GCT-covariant and VR-invariant Langevin equation is
equivalent to eq.(2.19).  We show that the
$\delta^n(0)$ divergences really cancel out.

With $\Phi_N $ substituted by means of constraint (4.2), action (4.1) becomes
  $$
   S_{N-1}=\int d^nxg^{\Ga \Gb}(\Phi)\partial_{\mu}
        \Phi_\Ga\partial^{\mu}\Phi_\Gb,\eqno(4.12)
       $$
       $$
           g^{\Ga \Gb}(\Phi)=\delta_{\Ga \Gb}+
       {\alpha\Phi_{\Ga} \Phi_{\Gb} \over 1-\alpha\Phi_{\Gg} \Phi_{\Gg}}
     ,\ \ \ \ \ \ \Ga,\Gb,\Gg = 1,2,\cdots,N-1.\eqno(4.13)
   $$
We can use Langevin equation (3.1) for the system,
because the assumption (3.21) is satisfied here
as shown in Appendix.
As we calculate the one-loop contributions to two-point functions,
we need Langevin equation to order $\alpha$, which is
 $$
    \dot\Phi_{\Ga} = \partial^2 \Phi_{\Ga} + \alpha \Phi_{\Ga}
        \partial_{\mu} \Phi_{\Gb} \partial_{\mu} \Phi_{\Gb} - \alpha
           (N-1) \delta^n(0) \Phi_{\Ga} + \alpha bN \delta^n(0)
         \Phi_{\Ga}
          + \xi_{\Ga} ,
        $$
      $$
     \xi_{\Ga} \equiv (\delta_{\Ga \Gb} - {\alpha\over 2} \Phi_{\Ga}
    \Phi_{\Gb}) \eta_{\Gb},\eqno(4.14)
 $$
where $\eta_{\Ga}$ is white noise.
We calculate two-point functions in one-loop order and show
that there remains no $\delta^n(0)$ divergence. Integral equation
corresponding to eq.(4.14) is
  $$
    \Phi_{\Ga}(x,t)=\int d \tau G(x,t-\tau)\MKO\eta_{\Ga}(x,\tau)
      -{\alpha \over 2} \Phi_{\Ga}(x,\tau) \Phi_{\Gb}(x,\tau)
       \eta_{\Gb}(x,\tau)
          $$
             $$
               +\alpha \Phi_{\Ga}(x,\tau)\partial_{\mu}
            \Phi_{\Gb}(x,\tau)\partial_{\mu}\Phi_{\Gb}(x,\tau) +
           \alpha\MCO bN-(N-1) \MCC \delta^4(0) \Phi_{\Ga}(x,\tau) \MKC
          ,\eqno(4.16)
         $$
where
  $$
      G(x,t)\equiv\int {d^4k \over (2\pi)^4}\ exp(-ikx)\theta(t)exp(-k^2t),
        \eqno(4.17)
   $$
which can be solved by iteration as eq.(4.8). Eq.(4.16) leads to the
following vertices
  $$
     Fig.\ 3
  $$
In Fig.3 we denote G by a line, $\eta_{\Ga}$ a cross or an
encircled cross and $\alpha\delta^n(0)$ a bullet, respectively.
The above vertices contribute to the propagator shown diagramatically in
Fig.4.
  $$
        Fig.\ 4
  $$
In one-loop order the $\delta^n(0)$-divergent contributions from
Figs.(4c), (4d) and (4e) are
  $$
    {-2\alpha N\over(k^2+{k^{\prime}}^2)^2}
      \theta(0)\delta_{\Ga \Gb}\delta^n(k+k^{\prime})\int d^np\
       ,\leqno(4c)
      $$
         $$
         {2\alpha(N-1)\over(k^2+{k^{\prime}}^2)^2}
        \delta_{\Ga \Gb}\delta^n(k+k^{\prime})\int d^np \ \ +\ \ 2nd\
        div.\
      ,\leqno(4d) $$
      $$
      {2\alpha(1-N+bN)\over(k^2+{k^{\prime}}^2)^2}(2\pi)^n\delta_{\Ga
    \Gb}\delta^n(k+k^{\prime})\delta^n(0)\ .\leqno(4e)
  $$
With $\theta(0)=b$, the sum of the above contributions is zero, which
coincides with the result of the constrained system. It seems that in
O(N) non-linear $\sigma$ model both Langevin equations (2.19) and (3.1)
lead to the correct results.

\section{5.Conclusion}

 We constructed Langevin equation (2.19) for constrained
system. In the derivation of (2.19), the improved treatment of the stochastic
consistency condition for constraints was essential. The Langevin equation
can be applied to any system obeying (2.1)-type constraint. From
the equation we directly derived Fokker-Planck equation and obtained
eq.(2.22) as the equilibrium distribution, which coincides with the one
obtained in path integral method. Owing to Ito calculus, the
Langevin equation contains $\delta^n(0)$-type singular terms and we
showed explicitly, in $O(N)$ non-linear $\sigma$ model, that the singular
terms are neccessary to the
cancellation of $\delta^n(0)$ divergences in one loop order.

 Furthermore, we ascertained that eq.(2.19) is equivalent to GCT-covariant
and VR-invariant Langevin equation (3.1). We applied eq.(3.1) to
O(N) non-linear $\sigma$ model and showed that the
singular terms in GCT-covariant and VR-invariant Langevin
equation are also necessary to the cancellation of $\delta^n(0)$ divergences
in one-loop order.

\section{Acknowledgments}
 We thank Prof.S.Kawasaki for careful reading of this manuscript and
also thank Dr.A.Nakamura for valuable discussion.

\vfill\eject

\section{Appendix}
 Here we show that assumption (3.21) is satisfied in O(N) non-linear
$\sigma$ model.
We choose polar coordinates as new variables $ Q^{\mu} $'s
in section 3 and immediately recognize that $e^A_{\ i} \nabla_B e^B_{\ i}$
vanishes. If the assumption (3.21) is ascertained in the above special
case, it is satisfied for any $\{ Q^{\mu}\}$ because of the covariance of
the assumption.

We start with a set of variables $\{ q_i \}$(i=1,2,$\cdots$,N)
obeying $ (q_i q_i)^{1\over 2} = r_c$($r_c$ = constant) and introduce
a new set of variables
$\{ Q^{\mu}\}$ ($\mu$=1,2,$\cdots$,N). In accordance with section 3
$$
   Q^a \equiv Q^N=(q_i\ q_i)^{1\over 2}-r_c.
$$
If we choose N-1 angles of N dimensional polar coordinates as
$ Q^{A} $'s ($A$=1,2,$\cdots$,N-1)
$$
\eqalign{
    q_1&=r_c\ \cos Q^1,\cr
    q_2&=r_c\ \sin Q^1\ \cos Q^2,\cr
       &\vdots\cr
q_{N-1}&=r_c\ \sin Q^1\ \sin Q^2\cdots \sin Q^{N-2}\ \cos Q^{N-1},\cr
    q_N&=r_c\ \sin Q^1\ \sin Q^2\cdots \sin Q^{N-2}\ \sin Q^{N-1},}
$$
we obtain the vielbein $e^{A}_{\ i}$
$$
\eqalign{
         &e^{A}_{\ i}\equiv {\partial Q^{A} \over \partial q_i}=
               {\cos Q^{A}\sin Q^{A +1}
                      \cdots \sin Q^{i-1}\cos Q^i
               \over r_c\ \sin Q^1  \cdots \sin Q^{A-1}},
                \ \ \ \ (i>A,i\not= N),\cr
         &e^{A}_{\ i}=0,\ \ \ \ \ \ \  (A > i),\cr
         &e^{A}_{\ i}={-\sin Q^{A} \over r_c\ \sin Q^1
                          \cdots \sin Q^{A-1}},\ \ \ \ (i=A),\cr
         &e^{A}_{\ N}=
                {\cos Q^{A}\sin Q^{A +1}
                          \cdots \sin Q^{N-1}
                \over r_c\ \sin Q^1  \cdots \sin Q^{A-1}},}
$$
and $e^a_{\ i}$
$$
\eqalign{  e^a_{\ i} &\equiv {\partial Q^N \over \partial q_i}
                 =\sin Q^1 \cdots \sin Q^{i-1} \cos Q^i,\ \ \ \ (i \not= N),\cr
           e^a_{\ N} &= \sin Q^1 \cdots \sin Q^{N-1}.}
$$
The vielbein satisfies (3.5)
$$
 e^{A}_{\ i} e^a_{\ i}=0.
$$
Vielbein $e^{A}_{\ i}$ leads to metric
$$
g^{A B} \equiv  e^{A}_{\ i} e^{B}_{\ i}
              ={1 \over r_c^2 \sin^2 Q^1 \cdots \sin^2 Q^{A-1}}
                            \delta^{A B}.
$$
By straightforward calculation we obtain
$$
 e^{A}_{\ i}\nabla_{B} e^{B}_{\ i}=0,
$$
which holds also for other choices of $\{ Q^{\mu} \}$
due to the covariance of the equation.
As we proved that assumption (3.21) is satisfied
in O(N) non-linear $\sigma$ model, we are allowed to use
the equation (3.1) in section 4.


\vfill\eject

\refout
\vfill\eject

\chapter{Figure Captions}
Fig.1  Vertices from eq.(4.8)

Fig.2  $\delta^n(0)$-divergent diagrams contributing to two-point function

Fig.3  Vertices from eq.(4.16)

Fig.4  Propagator including vertices shown in Fig.3

\end